\documentstyle[pra,two column,aps]{revtex}
\begin{document}
\title{ Quantum cryptography without public announcement bases }
\author{ Won Young Hwang \thanks{e-mail: hwang@chep6.kaist.ac.kr}
  and In Gyu Koh  }
\address{  Department of Physics,  Korea Advanced Institute of Science 
and  Technology, Yusung, Taejon, Korea }
\author{ Yeong Deok Han \thanks{e-mail: ydhan@core.woosuk.ac.kr}}
\address{ Department of Physics, Woosuk University, 
      Hujeong, Samrye, Wanju, Cheonbuk, Korea } 
\maketitle
\begin{abstract}
This paper provides a simple variation of the basic ideas of 
the BB84 quantum cryptographic scheme leading to a method of 
key expansion. A secure random sequence ( the bases sequence )
determines the encoding bases in a proposed scheme. 
Using the bases sequence repeatedly is proven to be safe by 
quantum mechanical laws.
\end{abstract}
\pacs{ 03.65.Bz 42.79.Sz 89.70+c } 

One of the most intriguing and exiting recent developments in 
quantum mechanics has been the prediction and demonstration of
a cryptographic key distribution scheme, the security of which 
is guaranteed by the laws of quantum mechanics \cite{wies}.
Theoretical models for the quantum key distributions ( quantum 
cryptography ) has been based on the uncertainty principle 
\cite{bene}, EPR states \cite{eker}, two nonorthogonal
states \cite{ben2} and Wheeler's delayed choice experiment
\cite{arde}.
In all the proposed quantum cryptographies \cite{wies}-\cite{arde},
there are {\it public announcement} steps  at which
Alice ( the sender ) and Bob ( the receiver ) exchanges some
informations on their operations via classical channel. 
Eve ( the eavesdropper ) has full access to the announced information 
on the classical channel but she can listen only and cannot 
tamper with the signals. 
In the first public discussion step of the standard BB84 scheme 
\cite{bene}, the bases on which Alice ( Bob ) encoded ( measured )
signals are announced each other.

 In this paper, a variation of
the BB84 scheme is proposed in which the public discussion 
of bases is not needed, although later public discussions
 for error detection and 'privacy amplification' 
\cite{ben3,hut2} is still necessary. 
Elimination of one public discussion step is an advantage 
by itself.  Furthermore, it reduces
information about bases to which Eve has access,  
which can be utilized in sophisticated eavesdropping strategy
\cite{deut}-\cite{cira}.

This paper is organized as follows. First, a simple form of 
quantum cryptography and it's weak point are presented.
 Next, it is described how this weak point is overcome in BB84
 scheme \cite{bene}.  After this introduction, 
a new method to overcome the weak 
point of the simple quantum cryptography is proposed. It is shown
that implementation of this method enables a quantum
cryptography without public announcement of bases.

Let us consider the following simple quantum cryptography. 
Alice sends to Bob some quantum carriers ( spin-$\frac{1}{2}$
particles or photons ) on which the 1-bit information 
is encoded; Alice encodes $ 0$ and $ 1$ on $|u+\rangle$ and 
$|u-\rangle$, respectively, where $|u+\rangle$ ( $|u-\rangle$ )
is the up ( down ) eigenstate of spin-measurement along axis 
$u$ which is known to only Alice and Bob. Since Alice and Bob use the
same  axis $u$, there is perfect correlation between the 
binary random sequence that Alice encoded 
and the one that Bob has retrieved by performing
spin-measurement along the axis $u$.
If Eve intercepts the particles sent to Bob by Alice 
and performs some spin-measurement on the particles and resends 
some particles to Bob later, then she inevitably
 will introduce some errors on 
the correlation between Alice's and Bob's.
 Eve can extract no information about the basis $u$
by any physical methods, because the mixture of $|u+\rangle$ and
$|u-\rangle$ with equal probabilities have a density operator
$\rho_u = \frac{1}{2}|u+\rangle \langle u+|      
       +\frac{1}{2}|u-\rangle \langle u-| = \frac{1}{2} I $
 which is identical to  
     $\rho_{u^\prime} =
 \frac{1}{2}|u^\prime+\rangle \langle u^\prime+| 
+ \frac{1}{2}|u^\prime-\rangle \langle u^\prime-|= \frac{1}{2} I $ 
, where $u \neq u^\prime$ and $I$ is the identity operator.
Thus, eavesdropping inevitably give rise to error which can
be detected by Alice and Bob,
 except for the case where
Eve knows the basis $u$ or Eve incidently measures along the 
$u$ basis. 

This simple quantum cryptography method has a few weak points.
First, there is considerable probability that Eve incidentally
 measures
along a basis almost similar to $u$ as there is only
one hidden parameter $u$. In that case, Eve obtains information
on the binary random sequence ( key ) introducing negligible errors.
Second, Eve might obtain information about the basis through
 indirect method:
she tries an arbitrarily chosen basis in eavesdropping. If the tried
basis is different from the basis $u$ which Alice and Bob use, then
errors are introduced which Alice and Bob will detect. Alice and Bob 
will discard the obtained data and start again. Now, Eve knows
that the chosen and tried basis is not the correct basis $u$.

The described above weak points of simple quantum cryptography
do not exist in BB84 scheme. They were eliminated in following
way. Alice uses randomly one of 
two bases $u$ and $u^\prime$ for encoding ( $u$ and 
$u^\prime$ are chosen $z$ and $x$ in BB84, respectively )
 and Bob performs spin-measurements along basis randomly 
chosen between $u$ and $u^\prime$. After all quantum carriers 
have arrived at Bob, Alice and Bob announce publically each 
other which  basis they chose at each instance. 
In about half of all the instances the  basis Alice chose are 
the same as those chosen by Bob. In these instances there will be
perfect correlation between Alice's and Bob's, unless the quantum 
carriers were perturbed by Eve or noise. In this way, using BB84
method, 
Alice and Bob can prevent Eve from knowing which  basis  Alice
choose to encode while Eve has access to quantum carriers.
Even if Eve knows about $u$ and 
$u^\prime$, Eve cannot learn anything about which of $u$ and $u^\prime$
is chosen by Alice at a particular instance before the public
announcement of bases.

 Public announcement of bases is a necessary step in
standard BB84 scheme. However, let us now consider 
 another shielding method, in which Alice and Bob 
do not publically reveal their bases. First, Alice and Bob share by 
any method ( by courier or by BB84 scheme )
 some secure binary random sequence that is known to nobody.  
 This random sequence is to be 
 used to determine the encoding  basis  $u$ and
$u^\prime$. Alice ( Bob ) encodes ( performs
spin-measurement ) on the  basis $z$ and $x$ when it is $0$
and $1$, respectively. For example, when the bases random sequence
is 
 \begin{equation}
  0,1,1,0,1.....
 \end{equation} 
and the signal random sequence that Alice wants to send to Bob is
 \begin{equation}
  1,0,1,0,1.....
 \end{equation}
 then she sends to him the following quantum carriers 
 \begin{equation}
  |z-\rangle, |x+\rangle, |x-\rangle, |z+\rangle, |x-\rangle,.....
 \end{equation}
Since Alice and Bob have common random sequence, 
there will be perfect correlation between them 
 unless the quantum carriers were perturbed by 
Eve or noise. Eve is naturally prevented from knowing 
about the encoding  bases, since she does not know the bases 
sequence. As we see, public announcement of bases is not 
needed in the proposed scheme. However, the scheme can 
only be useful if it is possible to use safely
the bases random sequence repeatedly. If this is not the case, 
 Alice and Bob have to consume the
same length of random sequences to obtain some length 
of new random sequences. Fortunately, quantum mechanical laws 
enable the bases random sequences to be used repeatedly enough, 
as shown below.

Suppose Alice used the bases random sequence of $N$ times
( $N$ is a positive integer ). 
 In order to know about the bases sequence Eve 
collects measurement records on the quantum carriers of all the 
$N$ times.
\footnote{In the sophisticated eavesdropping strategy
\cite{deut}-\cite{cira} the quantum carriers are not measured directly
and immediately. In those strategies, they make  the quantum
carriers to interact with auxiliary quantum states ( the ancilla ) and
they delay measurements on the ancilla by storing them 
in quantum memories
 until the time when they can extract maximal informations
about the key. In this case,
we deal with the ancilla stored in quantum memories instead of
records of measurement. However, we will get the same result 
that the two ensembles cannot be distinguished.}
 Next, she rearranges the records according to the order 
of the bases sequence. First, she collects the records of the first one
in each $ N$ sequences and label on this set number$ 1$,..., next
she collects the records of $ i$-th ($ 1 \leq i \leq N $, i is an integer )
one in all $N$
sequences and label on this set number $i$,...  Eve knows that
for each set of number $i$ , either $z$ or $x$ is used for encoding 
by Alice. Now, Eve tries to obtain some informations about which 
 basis is used for each set of number $i$. If it is possible,
Alice and Bob cannot use the bases sequence repeatedly. However, 
Eve can obtain no information about the  bases as shown below. 
When Alice 
encodes on $z$, Eve is given states as, for example, 
$|z-\rangle, |z+\rangle, |z+\rangle, |z-\rangle, |z-\rangle...$
with equal probabilities of $+$ and $-$. When Alice encodes on 
$x$, Eve is given states as, for example,
$|x+\rangle, |x+\rangle, |x-\rangle, |x+\rangle, |x-\rangle...$
with equal probability of $+$ and $-$. These two
 ensembles of states have the same density operator 
     $\frac{1}{2}|z+\rangle \langle z+|      
       +\frac{1}{2}|z-\rangle \langle z-|$ 
    ( $= \frac{1}{2}|x+\rangle \langle x+| 
        + \frac{1}{2}|x-\rangle \langle x-| $ ).
 Any two ensembles that have the same density operator give
statistically the
same outcome to any quantum mechanical measurements
\cite{hutt,bras}, even if they were composed of ensembles
of different state vectors \cite{d'es}. Suppose that Eve 
initially have some partial information ( in terms of Shannon's
information theory, the information  $I$ have a value between
 $0$ and $1$
in this case ) about the bases. Eve wants to increase
the information about the bases. However, since the two ensemble give 
rise to statistically the same outcomes to any quantum 
mechanical operations,
she cannot increase the information about the bases. Therefore we 
can assume that Eve's information about bases will remain zero if Eve's
initial one was zero. 
 It means that Eve cannot distinguish which basis is used and that 
the bases sequence can be used repeatedly.

It should be noted that the indistinguishability 
between ensembles of particles corresponding to
       $\frac{1}{2}|z+\rangle \langle z+|      
       +\frac{1}{2}|z-\rangle \langle z-|$ and
     $  \frac{1}{2}|x+\rangle \langle x+| 
      + \frac{1}{2}|x-\rangle \langle x-| $ 
can be shown 
 by another physical argument. If they are distinguishable we
can implement the superluminal communications using the
 spin-version \cite{bohm,sell} of Einstein-Podolsky-Rosen (EPR)
\cite{eins,sell} experiment. Let the state of source particle 
pairs is the singlet one
    $\frac{1}{\sqrt{2}}(|z+\rangle_1|z+\rangle_2
                       -|z-\rangle_1|z-\rangle_2)$
 (= $\frac{1}{\sqrt{2}}(|x+\rangle_1|x+\rangle_2
                       -|x-\rangle_1|x-\rangle_2)$ ).
If, at site $1$, one performs spin-measurement along $z$ ($x$) 
 axis, the states of particles given at site $2$ is equivalent
to $\frac{1}{2}|z+\rangle \langle z+|      
       +\frac{1}{2}|z-\rangle \langle z-|$ 
    ( $= \frac{1}{2}|x+\rangle \langle x+| 
        + \frac{1}{2}|x-\rangle \langle x-| $ ).
Thus if the one at site $2$ can distinguish between them, the one
at site $1$ can send signal instantaneously to the one at 
site $2$, by performing spin-measurement along $z$ or $x$
 axis according to the binary sequence he wants to send.

As the bases sequence can be used repeatedly 
the key can be distributed many times. However,
it should be underlined that the 
bases sequence have to be discarded after the expanded key 
is used for encrypting a message. It is because a cryptogram
gives partial information about the key by which it is encrypted.
With this information about the key Eve can extract information
about the bases. For example, if Eve know that $+$ ( or $0$)
is encoded on a quantum carrier she intercepted and the outcome
of spin-measurement along $x$ axis is $+$ ( or up state ),
then she knows that it is more probable that the basis is 
$x$ axis. Thus, after a cryptogram encrypted using the distributed
key is announced, it must be taken into account that Eve have 
considerable information about the bases sequence. One may be
concerned about the fact that this information on bases may 
be a weak point of 
the proposed scheme. However, this partial information
about bases is obtained after all the quantum 
carriers passed away Eve. Quantum cryptographic method 
successfully works as long as Eve does not know the encoding
bases while she has access to quantum carriers. In BB84 
scheme, indeed full information about bases is publically 
announced after quantum carriers passed away Eve. Taking
into account this fact we can argue
that the proposed scheme is not weaker than the BB84 scheme
in this respect.

 Similarly to BB84 scheme error check must be done in the  
proposed scheme, too.  Alice and Bob  compare some randomly
chosen subset of their key.  Bob inform publically 
to Alice whether he obtained $+$ or $-$  at the subset of
instances.  Next, Alice compares the 
informed data with her ones and check if there is error. 
  Public discussion  will be also needed in later
'privacy amplification' \cite{ben3,hut2}, to eliminate
discrepancies in bits which have not been revealed.  

In the proposed scheme, Alice and Bob prepare secure random sequence 
( which will be used as the bases sequence ) by  
courier  method or BB84 method, before distributing the key. 
As shown before, 
the bases sequence have to be discarded after
a cryptogram encrypted by the distributed key is announced. 
 It is inconvenient for Alice and Bob to  prepare the bases sequence 
everytime they distribute the key. It is possible to use more
convenient method: Alice and Bob leave some of
the distributed key in order to use it as basis sequence later.
 However, in this case, Alice and Bob must keep the remaining
key securely, what is a disadvantage of this method.

 We can conceive of the following indirect method for 
obtaining information about bases sequence.
 Eve can try various bases sequences, while listening to 
the public communication channel and observing the behavior
of Alice and Bob. If the tried bases sequence is incorrect one
( different from that of Alice and Bob ), error will be 
detected by Alice and Bob and they will discard the data.
Now, Eve, seeing that  they discard the data,
knows that the tried bases sequence 
is not a correct one. We can estimate that for Eve to 
learn bases sequence of $N$-bits, such indirect method should
be performed at an order of $2^N$ times. This fact means that 
the indirect method does not work for all practical purposes.
Furthermore, even the attack of the indirect method can be
avoided by delaying later steps until all quantum carriers
arrive at Bob. 

To summarize, we can say that 
this paper provides a simple variation of the basic ideas of 
the BB84 quantum cryptographic scheme leading to a method of 
key expansion. A secure random sequence ( the bases sequence )
determines the encoding bases in the proposed scheme. 
Using the bases sequence repeatedly is proven to be safe by 
quantum mechanical laws. There are three significant advantages of the
proposed method. First, public announcement of bases is not
needed. Second, it reduces information about bases to which 
Eve has access, which can be utilized in the sophisticated 
eavesdropping strategy. Third, there is no discarded data
 in ideal case, while 
in BB84 scheme about half of data is discarded. There is
also a disadvantage of the method as Alice and Bob must prepare 
a short secure random sequence to be used as a bases sequence. 
\section*{Acknowledgements}
We are very grateful to K.Chrzanowski
for helpful corrections of the paper.

\end{document}